\documentclass[10pt,journal,onecolumn]{IEEEtran}
\usepackage{dcolumn}% Align table columns on decimal point
\usepackage{bm}% bold math      
\usepackage{amsmath}
\usepackage{amssymb}
\usepackage{amsthm}
\usepackage{nccmath}
\usepackage{mathtools}
\usepackage{graphicx}
\usepackage{cite}
\usepackage{hyperref}
\usepackage{authblk}
\usepackage{epstopdf}
\usepackage{textcomp}
\usepackage{color}
\usepackage{tabularx}
\usepackage{lineno}

\newtheorem{definition}{Definition}
\newtheorem{proposition}{Proposition}

\newtheorem{corollary}{Corollary}

\begin{document}
\renewcommand*{\Affilfont}{\normalsize}
\setlength{\affilsep}{2em}   % set the space between author and affiliation
%\runningpagewiselinenumbers
%\linenumbers

\title{Normalised degree variance}
\author{
	Keith M. Smith$^{1,2,*}$, Javier Escudero$^{3}$
	\thanks{$^{1}$Centre for Medical Informatics, Usher Institute, University of Edinburgh, 9 Edinburgh Bioquarter, EH16 4UX, UK}
	\thanks{$^{2}$Health Data Research UK, Gibbs Building, Euston Rd, London, NW1 2BE, UK}
	\thanks{$^{3}$School of Engineering, Institute for Digital Communications, University of Edinburgh, West Mains Rd, Edinburgh, EH9 3FB, UK}
	\thanks{$^{*}$e-mail: k.smith@ed.ac.uk}%
}
\date{\today} 

%\IEEEtitleabstractindextext{%	

%}

\maketitle

\begin{abstract}
Finding graph indices which are unbiased to network size and density is of high importance both within a given field and across fields for enhancing comparability of modern network science studies. The degree variance is an important metric for characterising network degree heterogeneity. Here, we provide an analytically valid normalisation of degree variance to replace previous normalisations which are either invalid or not applicable to all networks. It is shown that this normalisation provides equal values for graphs and their complements; it is maximal in the star graph (and its complement); and its expected value is constant with respect to density for Erd\"os-R\'enyi (ER) random graphs of the same size. We strengthen these results with model observations in ER random graphs, random geometric graphs, scale-free networks, random hierarchy networks and resting-state brain networks, showing that the proposed normalisation is generally less affected by both network size and density than previous normalisation attempts. The closed form expression proposed also benefits from high computational efficiency and straightforward mathematical analysis. Analysis of 184 real-world binary networks across different disciplines shows that normalised degree variance is not correlated with average degree and is robust to node and edge subsampling. Comparisons across subdomains of biological networks reveals greater degree heterogeneity among brain connectomes and food webs than in protein interaction networks.
\end{abstract}

\section{Introduction}
The most fundamental parameters of a network are its number of nodes (network size) and number of edges (of which the ratio to total possible number of edges is the network density). The richness of topologies elicited by these simple building blocks has fascinated mathematicians for centuries and science, in general, for decades since the explosion of interest in understanding real-world complex networks. Comparing networks of different sizes and/or densities is difficult because of the dependency of many network features on these fundamental parameters \cite{Dormann2009}, yet the ability to compare between different sizes and densities is necessary to gain an unbiased understanding of the topologies in a given field as well as across fields. Indeed, the fact that differently sized networks explain similar phenomena in a given area is commonplace. For example, social networks depend on the number of participants in the study \cite{VegaRedondo2007}; infrastructural networks within cities will vary in size according to the size of the city \cite{Batty2008}; and networks constructed from sensors, such as in electrophysiology, depend on the number of sensors used \cite{Smith2017b}.

One of the most widespread forms of analyses of real-world networks is that of defining some topological measurement from which one can compare the networks against other networks and null models in order to gain an understanding of the networks' particular characteristics \cite{Newman2010}. For example, in brain networks, network measures can help inform of brain connectivity differences between patients with some condition and healthy controls \cite{Bullmore2009}. The degree distributions of complex networks have been long been an important area of study, where real-world networks generally exhibit heavy tails \cite{Strogatz2001}. The heterogeneity of degrees has been an active area of study in complex networks for some time. The consistent feature of heterogeneity measures is that they are generally maximal in star-like graphs and minimal in regular graphs (where all nodes have same degree). In this way, they measure the inequality of the distribution of the number of connections in the network. The degree variance is the original characteristic \cite{Snijders1981}. Several other measures of heterogeneity have been proposed \cite{Bell1992, Jacob2017, Safei2019}, but the degree variance remains one of the most widespread and easiest to interpret, aiding, for example, in analysing degree distributions of networks from genetics \cite{Horvath2008}, physics \cite{Cao2015}, chemistry \cite{Randic2016}, sociology \cite{Phillips2019}, neuroscience \cite{Smith2017b}, and policy \cite{Mochtak2019, Makinen2019}, among others. 

%The degree distribution of a network is a key property for assessing its topology \cite{Newman2010}.  One important characteristic of the degree distribution is the degree variance \cite{Snijders1981}, $v(G) = var(\mathbf{k})$, which can aid in understanding the relationship of degree hierarchies and hub dominance \cite{Smith2017}. 

However, variance depends on the size of samples used, which varies accordingly for fixed density and varying size-- larger size needing more edges to account for the same density. Similarly, the degree distribution is dependent on the number of edges in the network since the maximum possible degree is fixed by network size.  Thus, we need to work towards effective and efficient control of the scaling of network size and density in the measurement of degree variance. This is of particular necessity to support increasingly critical interdisciplinary areas using large multi-dimensional datasets.

Here, we propose a mathematically rigorous and computationally efficient normalisation of degree variance, $\bar{v}$, with a closed form expression. Particularly, we will prove that $\bar{v}\in[0,1]$ for all graphs. Not only this, we will show that graph complements achieve equivalent values of $\bar{v}$; that $\bar{v}$ only satisfies unity asymptotically for the star graph (and its complement) as $n\rightarrow \infty$; and that $\bar{v}$ is independent of network density for Erd\"os-R\'enyi (ER) random graphs. Furthermore, our normalisation is also well defined for graphs with isolated nodes and thus, we argue, of broader scope than the other relevant normalised measure of heterogeneity \cite{Estrada2010}. Demonstrations are provided that our normalisation is the least variable normalisation with respect to network size and density for a number of network types and that it is also computationally efficient. Finally, we show an application of the new normalisation to 184 real-world networks and its robustness to node and edge subsampling of these networks.

\section{Background}
A simple network is defined by a set of nodes, $\mathcal{V}= \{1,\dots,n\}$, connected together by a set of edges, $\mathcal{E}= \{(i,j): i,j\in\mathcal{V}\}$. Network size is then $|\mathcal{V}| = n$. The convention is that $|\mathcal{E}| = 2m$ including each of the $m$ edges twice ($(i,j)$ \& $(j,i)$). The largest possible size of $\mathcal{E}$ is obtained in the complete graph with $2m = n(n-1)$. Thus, network density is $ d = 2m/n(n-1)$. The degree of a node, $k_{i}$, is simply the number of edges adjacent to it, and we denote the set of degrees of a graph as $\mathbf{k} = \{k_{1},\dots,k_{n}\}$. For reference, all repeated notations in the article are in Table \ref{notate}.

\begin{table}[!tb]
	\centering
	\caption{\label{notate} Notation}
	\begin{tabular}{l|l|l|l}
			\multicolumn{2}{l|}{\textbf{Basic network notation}} &\multicolumn{2}{l}{\textbf{Network heterogeneity notation}}\\
			\hline
			$n$ 					&Number of nodes 				&$G_{qs}$ 		&Quasi-star graph				\\
			$m$ 					&Number of edges 				&$G^{*}$	 		&Perfect quasi-star graph			\\
			$d$ 					&Network density				&$v$ 			&Degree variance				\\
			$\mathcal{V}$ 			&Node set						&$\bar{v}$ 		&Normalised degree variance		\\
			$\mathcal{E}$ 			&Edge set 					&$J$ 			&Quasi-star normalisation	 of		\\
			$G$ 					&Graph						&				&degree variance				\\	
			$k_{i}$ 				&Degree of node $i$ 			&$\sigma^{2}$ 		&Average degree normalisation	 of	\\
			$\mathbf{k}$ 			&Degree set  					&				&degree variance				\\
			$\hat{G}$ 				&Complement of $G$			&$\rho$ 			&Heterogeneity index			\\
			$\hat{m}$ 				&Number of edges in $\hat{G}$ 	&$r$ 			&Number of dominant nodes in		\\
			$\hat{k}_{i}$ 			&Degree of node $i$ in $\hat{G}$ 	&				&quasi-star graph				\\
			$p$					&Edge probability in				&				&							\\
								&random graph					&				&							\\	
	\end{tabular}
\end{table}

%\begin{table*}[!tbh]
%	\centering
%	\caption{\label{notate} Notation}
%	\begin{tabular}{c|l|c|l|c|l}
%			$n$ 			&number of nodes 		&$\hat{G}$ 		&complement of $G$			&$\rho$ 			&heterogeneity index\\ 
%			$m$ 			&number of edges 		&$\hat{m}$ 		&number of edges in $\hat{G}$ 		&$G^{*}$ 			&perfect quasi-star graph\\
%			$d$ 			&network density		&$\hat{k}_{i}$ 		&degree of node $i$ in $\hat{G}$ 	&$r_{s}$ 			&Spearman correlation coefficient \\
%			$\mathcal{V}$ 	&node set				&$v$ 			&degree variance				&$med$ 			&median\\
%			$\mathcal{E}$ 	&edge set 			&$\bar{v}$ 		&normalised degree variance		&$\bar{v}_{sub}$ 	&$\bar{v}$ of subsampled network\\
%			$G$ 			&graph				&$G_{qs}$ 		&quasi-star graph				&$\rho_{sub}$ 	&$\rho$ of subsampled network\\
%			$k_{i}$ 		&degree of node $i$ 		&$J$ 			&quasi-star normalisation			&				&\\
%			$\mathbf{k}$ 	&degree set  			&$\sigma^{2}$ 		&average degree normalisation		& 				&\\
%	\end{tabular}
%\end{table*}

Degree variance is seen as a measure of graph heterogeneity which is essentially conceptually equivalent to graph irregularity. Regular graphs have been of interest to mathematicians for many years, at least since the pioneering work of Petersen in 1891 \cite{Petersen1891}. However, it was not until the seminal paper of Collatz and Sinogowitz in 1957 \cite{Collatz1957} that indices to characterise the irregularity of graphs became a topic of interest. In \cite{Collatz1957} it was proposed to study the difference between the largest eigenvalue of the graph adjacency matrix and the average degree, $\epsilon(G) = \lambda_{1}(G)-\sum_{i}k_{i}/n$, claiming that this measure was only zero in connected graphs for regular graphs. It was conjectured that star graphs attained the highest value of $\epsilon(G)$, but this was disproved by Cvetkovi\'c \& Rowlinson \cite{Cvetkovic1988}.

With the booming interest in applications of graph theory to real-world networks, the degree variance, $v(G) = var(\mathbf{k})$, was proposed as a measure of graph heterogeneity in 1981 \cite{Snijders1981}. Later, Bell \cite{Bell1992} compared $v(G)$ and $\epsilon(G)$ and found incompatibilities between them, noting that the measures had different relative values for certain pairs of graphs. Using the constructs of quasi-complete and quasi-star graphs introduced by Ahlsewade \& Katona \cite{Ahlsewade1978}, Bell then proved that $\epsilon(G)$ was in fact maximal only for quasi-complete graphs, whereas $v(G)$ was maximal either for a quasi-complete or a quasi-star graph for any given network size and density (although not necessarily uniquely) \cite{Bell1992}.

A quasi-star graph is a graph consisting of $r$ dominant nodes and at most one node of degree $s>r+1$ connected to other non-dominant nodes, here referred to as a developing node, for a graph with $r(n-1) + s$ edges (Fig 1, left). If $s = 1$ then the remaining edge connects two non-dominant nodes so that they both have degree $r+1$. A quasi-complete graph is a graph consisting of a complete subgraph of order $s$ and at least $n-s-1$ isolated nodes, with the remaining node connected to $r$ nodes of the complete subgraph for a graph with $s(s-1) + r$ edges, see Fig 1, right. We regard here as perfect quasi-complete and quasi-star graphs as such graphs without any remainder, $s$. These are special in that the perfect quasi-complete graph of order $r$ is the complement of the perfect quasi-star graph of order $n-r$. 

\begin{figure*}[!tbh]
	\centering
	\includegraphics[trim= 70 50 0 80,clip,scale = 0.48]{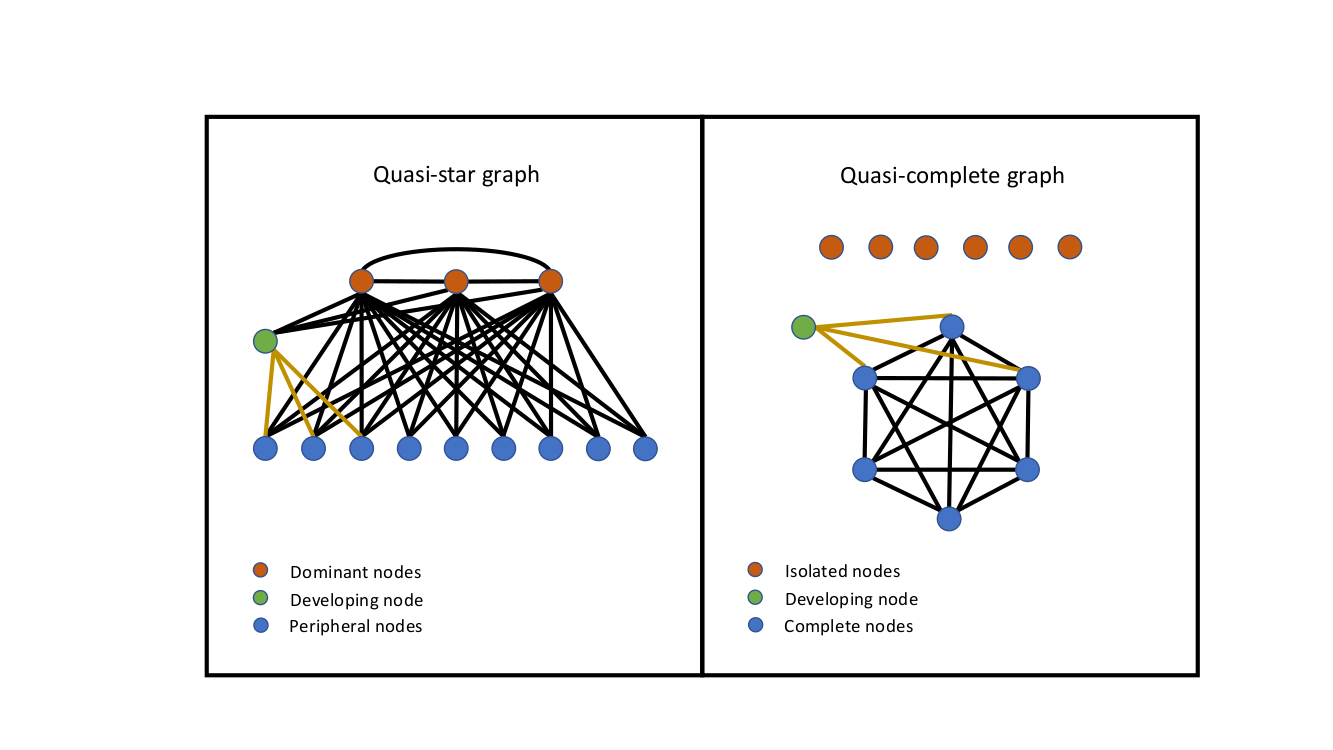}
	\caption{Illustration of a quasi-star graph, left, and a quasi-complete graph, right. The golden edges refer to the edges making up the remainders in the graph construction process for a specified density}
	\label{Quasi}
\end{figure*}

In a related topic, Abrego \emph{et al.} \cite{Abrego2008} proved which of either the quasi-star or quasi-complete graphs with $n$ nodes and $m$ edges obtained the maximum sum of squares of degrees. It was then shown by Smith \& Escudero \cite{Smith2017} that maximising the sum of squares of degrees for fixed $n$ and $m$ was equivalent to maximising the degree variance by the equation
\begin{equation}\label{eq1}
v(G) = \frac{||\mathbf{k}||^{2}_{2}}{n-1} - \frac{(2m)^{2}}{n(n-1)},
\end{equation}
where $||.||_{2}$ is the $l_{2}$-norm, and thus Abrego \emph{et al.}'s result was also seen to hold for $v(G)$.

In the work by Snijders \cite{Snijders1981}, a normalisation of the degree variance was attempted through division by the value achieved for the quasi-star graph with same size and density because it was thought that the quasi-star graph was always maximal. For the quasi-star graph with same size and density as a graph $G$, $G_{qs}$, this can be written
\begin{equation}
J = \frac{v(G)}{v(G_{qs})}.
\end{equation}
But the previously noted results render such a normalisation invalid. Apart from this, one needs to construct the quasi-star graph to compute it, rendering a general mathematical analysis of this property infeasible and causing computational efficiency problems for very large graphs. 

Normalisation of the degree variance has also been proposed through division by the average degree, $\left<k\right>$ of the network \cite{Zimmerman2004},
\begin{equation}
\sigma^{2}(G) = \frac{v(g)}{\left<k\right>}.
\end{equation}
However the average degree depends on network density and thus network size. 

The other notable class of measure of graph irregularity are those based on absolute differences of degrees. This was originated by Albertson \cite{Albertson1997} with the measure 
\begin{equation}
irr(G) = 1/2\sum_{i,j}|k_{i}-k_{j}|.
\end{equation}
Estrada \cite{Estrada2010} then provided a normalised measure based instead on inverse square roots of degrees:
\begin{equation}\label{eq2}
\rho(G) = \frac{\sum_{ij}\left(k^{-1/2}_{i}-k^{-1/2}_{j}\right)^{2}}{n- 2\sqrt{n-1}},
\end{equation}
by relating it to the Randi\'c index and using its known upper bounds \cite{Li2008}. Estrada argued that this measure was maximised only by star graphs and not quasi-complete graphs. However, in actuality the measure is not well defined for any graph with isolated nodes, such as quasi-complete graphs, since this leads to division by zero in the terms of $k_{i}^{-1/2}$ for isolated $i$. On top of this, as we shall see, this measure is biased to network density where graphs with larger densities can be expected to obtain lower values than graphs with small densities.

\begin{table*}[!tbh]
	\centering
	\caption{Traits of degree heterogeneity among proposed normalisations}\label{survey}
	\begin{tabular}{c|l|l|l|l|l|l|l}
					&      				&\textbf{Defined for}		&\textbf{Computational}	&\textbf{Complements} 		&\textbf{Value for}	&\textbf{Value for} 		&\textbf{Value for}\\
		\textbf{Metric}	&\textbf{Normalised}	&\textbf{all networks}		&\textbf{efficiency} 		&\textbf{have equal values} 	&\textbf{star graph}	&\textbf{regular graph} 	&\textbf{random graph}\\
			\hline
		$v$			&No				&Yes					&High				&Yes						&$\rightarrow \infty$ as $n\rightarrow \infty$	&0		&$(n-1)p(1-p)$\\		
			\hline
			$\bar{v}$		&Yes		&Yes		&High		&Yes		&$\rightarrow 1$ as $n\rightarrow \infty$		&0		&$\approx2/n$\\			
			
			$\sigma^{2}$\cite{Zimmerman2004}	&No		&Yes		&High		&No		&$\rightarrow \infty$ as $n\rightarrow \infty$	&0		&depends on $n$ and $p$\\	
			$J$\cite{Snijders1981}			&No		&Yes		&Low		&No		&1									&0		&depends on $n$ and $p$\\	
			$\rho$\cite{Estrada2010}	&Yes		&No		&Moderate	&No		&1									&0		&depends on $n$ and $p$\\	
	\end{tabular}
\end{table*}

Thus, a normalised heterogeneity index which is applicable to all graph types and is invariant to network size and density is of interest to help resolve these outstanding issues. Table \ref{survey} provides a comparison of traits for normalised measures of heterogeneity, for reference, which is collected on known information from the relevant cited literature and from the observations of our experiments in section IV.

\section{Normalised Degree Variance}
We will begin with the details of the proposal of our normalisation of degree variance.
\begin{proposition}
For any graph $G$ with density $d = \frac{2m}{n(n-1)}\in(0,1)$, the identity
\begin{equation}\label{mainresult}
\bar{v}(G) = \frac{n-1}{nm(1-d)}v(G)
\end{equation}
is bounded in the interval $[0,1]$.
\begin{proof}
From \eqref{eq1}, normalisation for degree variance can be achieved given a known upper bound of $||\mathbf{k}||^{2}_{2}$ for a graph. Fortunately, this exists from a result by de Caen \cite{deCaen1998}:
\begin{equation}
||\mathbf{k}||^{2}_{2} \leq m\left(\frac{2m}{n-1} + n - 2\right).
\end{equation}
Substituting this into \eqref{eq1}, we have
\begingroup
\addtolength{\jot}{0.75em}
\begin{align}
v(G) &\leq \frac{m\left(\frac{2m}{n-1} + n - 2\right)}{n-1} - \frac{(2m)^{2}}{n(n-1)}\\
\iff \frac{n-1}{m}v(G)&\leq \frac{2m}{n-1} + n - 2 - \frac{4m}{n}\\
\iff \frac{n-1}{nm}v(G) %&\medmath{\leq  \frac{2m}{n(n-1)} + 1 - \frac{2}{n} - \frac{4m}{n^{2}}}\\
&\leq 1 - \frac{2}{n} + \frac{4m - 2mn}{n^{2}(n-1)}\\
&\leq^{*} 1 - \frac{2}{n} + \frac{2n(n-1)-2mn}{n^{2}(n-1)}\\
&= 1 - \frac{2m}{n(n-1)}\\
&= 1-d\\
\iff \bar{v}(G) & \leq 1,
\end{align}
\endgroup
where $^{*}$ comes from the fact that $2m\leq n(n-1)$.
\end{proof}
\end{proposition}
Now, in the special cases that $d = 0$ or $1$, we obtain the empty and complete graphs, respectively, and the denominator of \eqref{mainresult} would be 0, meaning that the function would be undefined (given that a graph can only be defined for $n>0$). But empty and complete graphs are regular graphs with (non-normalised) degree variance 0. Indeed, one definitely agreed characteristic of heterogeneity is that regular graphs obtain zero values. Thus, to overcome this we simply define normalised degree variance at $d = 0$ and $1$ to be 0. This provides us with the following definition:
\begin{definition} For a graph, $G$, with $n$ nodes, $m$ edges and density $d = 2m/n(n-1)$, its normalised degree variance, $\bar{v}(G)$, is
\begin{align}
\bar{v}(G) 		& = \frac{n-1}{nm(1-d)}v(G), 	&d \in(0,1)\\
\bar{v}(G) 		& = v(G) = 0,	& d = 0 \text{ or } 1.
\end{align}
\end{definition}

%with respect to quasi-star and quasi-complete graphs remain to be seen. 
Note that the matter of complete and empty graphs is a purely technical point. In almost all practical applications networks exhibit non-trivial topologies with $d\in(0,1)$. Now, clearly $\bar{v}$ attains its lower bound, 0, for all regular graphs (where $v(G) = 0$ and given previous discussion of empty and complete graphs). It remains to assess how good is the upper bound of 1. Further, we must check $\bar{v}$'s behaviour with respect to network size and density. The following results prove that $\bar{v}(G)$ for a star graph of size $n$ tends to 1 as $n\rightarrow\infty$, and that perfect quasi-star graphs are a decreasing function of density, thus the star graph is always maximal amongst them. These results validate degree variance as a relevant measure of heterogeneity and, as far as we are aware, the only normalised measure of heterogeneity that is valid for all simple graphs, including those with isolated nodes.

\begin{corollary}
Let $G^{*}(n,r)$ be the perfect quasi-star graph with $n$ nodes and $r$ dominant nodes, then the following statements hold:
\begingroup
\addtolength{\jot}{0.75em}
\begin{enumerate}
	\item $\bar{v}(G^{*}(n,1)) \rightarrow 1$ as $n\rightarrow\infty$
	\item $\bar{v}(G^{*}(n,r))$ is a monotonically decreasing function with respect to $r$ for $n\geq1, r\geq 0\in\mathbb{R}$
\end{enumerate}
\endgroup
\begin{proof}
\begin{enumerate}
\item For a perfect quasi-star graph, $m$ is made up of $r$ dominant nodes of degree $n-1$ and $n-r$ nodes of degree $r$. The general degree sequence of a perfect quasi-star graph is thus
\begin{equation}
\{r,\dots,r,n-1,\dots,n-1\},
\end{equation}
and
\begingroup
\addtolength{\jot}{0.75em}
\begin{align*}
2m  &= (n-r)r + r(n-1) \\
	&= r(2n-r-1),
\end{align*}
\endgroup
so that
\begingroup
\addtolength{\jot}{0.75em}
\begin{align}
1- d 	&= 1-\frac{r(2n-r-1)}{n(n-1)}\\
	&= \frac{n(n-1)-r(2n-r-1)}{n(n-1)}\\
	& = \frac{n^{2} + r^{2} -2rn -(n-r)}{n(n-1)}\\
	& = \frac{(n-r)(n-r-1)}{n(n-1)}.
\end{align}
\endgroup
Then
\begingroup
\addtolength{\jot}{0.75em}
\begin{align}
\bar{v}(G^{*}(n,r)) &= \frac{1}{1-d}\left(\frac{(n-r)r^2 + r(n-1)^2}{nm} -\frac{(2m)^2}{n^2m}\right),\\ 
	&=\frac{2(n-1)(r(n-r)n + (n-1)^{2}n- r(2n-r-1)^{2})}{n(n-r)(n-r-1)(2n-r-1)}, \label{numden}
%				&\medmath{=\tfrac{n(n-1)\splitfrac{2(r^{2}(n-r)n + r(n-1)^{2}n} {- r^{2}(2n-r-1)^{2})}}{(n-r)(n-r-1)r(2n-r-1)n^{2}}}, \label{numden}
\end{align}
\endgroup
and for a the star graph with $r = 1$, this simplifies to
\begingroup
\addtolength{\jot}{0.75em}
\begin{align}
\bar{v}(G^{*}(n,1)) &= \frac{n^{2}(n-1)- (2n-2)^{2}}{n(n-2)(n-1)}\\
				&= \frac{n^{3} - 5n^{2} + 8n - 4}{n^{3}-3n^{2}+2n} \rightarrow 1 \text{ as } n\rightarrow\infty,
\end{align}
\endgroup
as required.
\item It is well known that a function is monotonically decreasing if and only if its derivative is always less than or equal to zero. Thus, for $f(r) = \bar{v}(G^{*}(n,r))$, we prove that $f'(r)\leq 0\  \forall n\geq r\geq 0$.

First we factorise the numerator of \eqref{numden}. It is straightforward to compute for $r=n$ and see that this is a root of this polynomial. Then, by long polynomial division we find that $r = n-1$ is a double root of this polynomial and the numerator factorises to give
\begingroup
\addtolength{\jot}{0.75em}
\begin{align}			
\bar{v}(G^{*}(n,r))		&= \frac{2n(n-1)(n-r)(n-r-1)^{2}}{(n-r)(n-r-1)(2n-r-1)n^{2}}.\label{starfactorised}\\
					&=\frac{2(n-1)(n-r-1)}{(2n-r-1)n}
\end{align}
\endgroup
We now differentiate with respect to $r$ using the quotient rule to get
\begingroup
\addtolength{\jot}{0.75em}
\begin{align}
\frac{\partial}{\partial r}\bar{v}(G^{*}(n,r)) &=\frac{-2(n-1)(2n-r-1)n+n2(n-1)(n-r-1)}{(2n-r-1)^{2}n^{2}}\\
%&\medmath{=\frac{-2(n-1)n}{(2n-r-1)^{2}n}}\\
&= \frac{-2(n-1)}{(2n-r-1)^{2}} \leq 0 \ \forall n\geq 1, r\geq0 \in\mathbb{R}
\end{align}
\endgroup
as required.
\end{enumerate}
\end{proof}
\end{corollary}

This shows that 1 is indeed a tight upper bound of $\bar{v}$, that this is approached by the star graph and its complement and that there is a decreasing tendency with respect to increasing density for quasi-star graphs. In Appendix I, we conduct mathematical analysis to explore lower bounds for a network with a given degree distribution. In fact, we show that a lower bound can be achieved knowing a specific fraction of nodes having a certain degree, and a maximum lower bound from this can be found applying the formula to all degrees of the network and choosing the maximum result. Of note, generally the maximum lower bound is found at either the largest or smallest degree in the network, highlighting the relationship between degree variance and the network's inequality of degrees.

Now, while the degree variance gives equivalent values for graphs and their complements, it is clear that the heterogeneity index does not, since the value of a star graph is $1$ while the value of its complement, the quasi-complete graph of order $n-1$, is undefined. We will now show that the proposed normalisation of degree variance also provides equivalent values for graphs and their complements.

\begin{proposition}
For a graph $G$ with complement $\hat{G}$,
\begin{equation}
\bar{v}(G) = \bar{v}(\hat{G}).
\end{equation}
\begin{proof}
For a graph $G$ with $0<m<n(n-1)$ edges (i.e. non-empty and non-complete) and degrees $k_{1},\dots,k_{n}$ so that $\sum_{i=1}^{n}k_{i} = 2m$, its complement, $\hat{G}$, has $\hat{m} = (n(n-1)-2m)/2$ edges and degrees $\hat{k}_{i} = n-1-k_{i}$. Then
\begin{align}
\bar{v}(G) &= \frac{n(n-1)\sum_{i=1}^{n}k_{i}^{2} - (n-1)(2m)^{2}}{nm(n(n-1)-2m)},
\end{align}
and
\begingroup
\addtolength{\jot}{0.75em}
\begin{align}
\bar{v}(\hat{G}) &=\frac{n(n-1)\sum_{i=1}^{n}\hat{k}_{i}^{2} - (n-1)(2\hat{m})^{2}}{n\hat{m}(n(n-1)-2\hat{m})}\\
		&= \frac{n(n-1)\sum_{i=1}^{n}(n-1-k_{i})^{2} - (n-1)(n(n-1)-2m)^{2}}{n2m(n(n-1)-2m)/2}\\
		&=\frac{n(n-1)\left(n(n-1)^{2} + \sum k_{i}^{2} - 2(n-1)\sum k_{i}\right) - (n-1)(n^{2}(n-1)^{2}-4n(n-1)m + (2m)^{2})}{nm(n(n-1)-2m)}\\
		&=\frac{n^{2}(n-1)^{3} + n(n-1)\sum k_{i}^{2} - 4n(n-1)^2m- n^{2}(n-1)^{3} + 4n(n-1)^{2}m - (n-1)(2m)^{2}}{nm(n(n-1)-2m)}\\
		&= \frac{n(n-1)\sum k_{i}^{2} - (n-1)(2m)^{2}}{nm(n(n-1)-2m)} \\
		&= \bar{v}(G),
\end{align}
\endgroup
as required. Finally, the empty graph ($m=0$) is the complement of the complete graph ($m = n(n-1)$), both having $\bar{v}(G) = 0$.
\end{proof}
\end{proposition}

\subsection{Normalised degree variance of ER random graphs}
The simple and well-known statistical properties of ER random graphs allows us to calculate $\bar{v}$ for ER random graph ensembles directly. Note, there are two versions of ER random graph ensembles. $G(n,p)$ employs $p$ as the uniform probability of the existence of edges. Subsequently, because the calculation of the existence of each of the $n(n-1)/2$ possible edges is equivalent to a Bernoulli trial with probability $p$, the number of edges in a realisation of $G(n,p)$ varies and follows a binomial distribution $m\sim B(n(n-1)/2,p)$ \cite{Erdos1959}. The other prevalent version of the ER random graph ensemble, $G(n,m)$, allows for the control of the number of edges by generating numbers uniformly at random from $[0,1]$ for each possible edge and choosing those possible edges with the $m$ largest values as existent \cite{Gilbert1959}. We shall focus here on $G(n,m)$ since we wish to control for number of edges in our experiments. The equivalent proof for $G(n,p)$ is more involved and is provided in Appendix I.

\begin{corollary}\label{Gnm}
For the ER random graph ensemble $G(n,m)$
\begin{equation}
	\bar{v}(G(n,m)) = \frac{2(n-1)^2}{(n^3+n^2)} \rightarrow 0 \text{ as } n\rightarrow\infty \ \forall m,
\end{equation}
and so does not depend on $m$.
\begin{proof}
From a previous result from Gutman and Paule \cite{Gutman2002}, we have:
\begingroup
\addtolength{\jot}{0.75em}
\begin{align}
v(G(n,m))  &= \frac{2m(n^2-n-2m)}{n^3+n^2} \\
		&= \frac{2m(n(n-1)-2m)}{n^3+n^2} \\
		&= \frac{2mn(n-1)\left(1-\frac{2m}{n(n-1)}\right)}{n^3+n^2} \\
		&= \frac{2m(n-1)(1-d)}{n^2+n}.
\end{align}
\endgroup
Using our normalisation we obtain
\begingroup
\addtolength{\jot}{1em}
\begin{align}\label{randnorm}
\bar{v}(G(n,m)) 	&= \frac{n-1}{nm(1-d)}\frac{2m(n-1)(1-d)}{n^2+n}\\
			&= \frac{2(n-1)^2}{n^3+n^2}\rightarrow 0\text{ as } n\rightarrow\infty,
\end{align}
\endgroup
as required.
\end{proof}
\end{corollary}
Importantly, it is clear from this result that $\bar{v}(G)$ is unbiased to network density for ER random graphs $G(n,m)$. Furthermore, it decays fairly slowly towards $0$ at a rate of $1/n$. %Note that this result holds in all cases that $G(n,q)$ has a binomial degree distribution (the vast majority of cases), but does not hold, for example, in the redundant cases that $q = 1$ or $q = 0$ giving complete and empty graphs, respectively.

\section{Experiments}
\subsection{Validity and stability with respect to network size of normalisation approaches}
Counter examples for the previously proposed normalisations are not difficult to come by. To demonstrate this we shall consider values for quasi-star graphs, quasi-complete graphs, weighted ER random graphs \cite{Erdos1959}, weighted random geometric graphs \cite{Dall2002}, scale-free networks \cite{Barabasi1999}, random hierarchy models \cite{Smith2017} and Electroencephalogram (EEG) networks of size $n=16,32,64$ and $128$. 

For the ER random graphs and random geometric graphs, 100 realisations were generated for each network size and results averaged. The scale-free networks were chosen at three densities for each network size. Scale-free graphs of density 12.5\% were obtained by selecting 12.5\% of nodes as the core subgraph and the average degree of additional nodes as $6.25\%n$. Graphs of density roughly 25\% were obtained by selecting 25\% of nodes as the core subgraph and the average degree of additional nodes as $12.5\%n$. Graphs of density roughly 37.5\% were obtained by selecting 37.5\% of nodes as the core subgraph and the average degree of additional nodes as $25\%n$. For each, 100 networks were generated and results averaged.

Random hierarchy models generate networks with hierarchical degree structure determined by three parameters: the number of hierarchical levels, the strength of separation of levels and the probability that a node exists in a given level \cite{Smith2017}. In our case, the strength parameter was randomly selected between 0.05 and 0.5-- in steps of 0.05-- and with hierarchical levels randomly selected from 2 to 5, while the level probability distribution was fixed as a geometric distribution with probability 0.6. For each, 1000 networks were generated (due to greater variability of topologies than other models) and results averaged. 

The EEG networks are derived from a 129 node EEG eyes open dataset.  This is available online under an Open Database License via the Neurophysiological Biomarker Toolbox tutorial \cite{NBT}. It consists of data for 16 volunteers. We have previously used the data for which full processing details can be found in \cite{Smith2017b}. Weighted connectivity adjacency matrices were computed using the phase-lag index (PLI) \cite{Stam2007}. To get a network of size $n$ from these EEG networks, $n$ electrodes were chosen at random 100 times and results averaged.

\begin{figure*}[!t]
	\centering
	\includegraphics[trim= 130 30 0 40,clip,scale = 0.19]{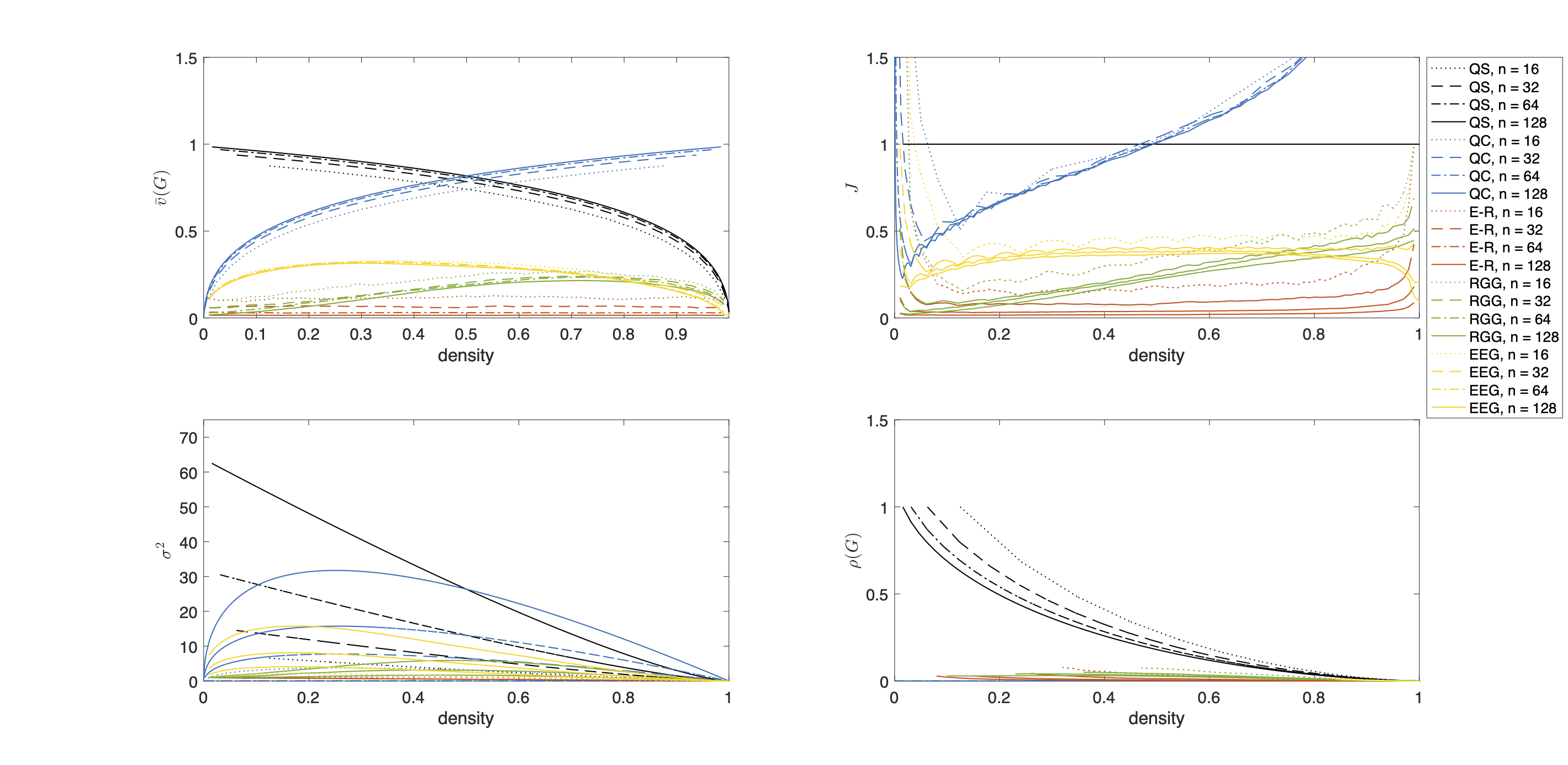}
	\caption{Normalisations of degree variance of quasi-star graphs (QS), quasi-complete graphs (QC), ER random graphs (E-R), Random Geometric Graphs (RGG) and EEG PLI networks (EEG) of sizes 16 (dotted lines), 32 (dashed lines), 64 (dash-dotted lines) and 128 (solid lines). The proposed normalisation shows smooth curves which are markedly stable with respect to $n$ for all graphs. Clearly, $J$ and $\sigma^2$ violate standard normalisation principles and are influenced by network size. On the other hand $\rho$ cannot be computed for graphs with isolated nodes which are particular prominent at lower densities of the weighted models and values are squeezed at high densities.}
	\label{fig1}
\end{figure*}

For each $n$, the perfect quasi-star graphs and perfect quasi complete graphs for each $d$, and integer percentage binarisation thresholds of ER random graphs, random geometric graphs, random hierarchy models and EEG networks are computed and we take the proposed normalisation, $\bar{v}$, quasi-star normalisation, $J$, average degree normalisation, $\sigma^2$, as well as the heterogeneity index, $\rho$, of these networks. Fig. \ref{fig1} shows the obtained values plotted against density.

The proposed normalisation works as expected with all values in $[0,1]$. The maximum value is achieved by the largest ($n=128$) star graph (perfect quasi-star with $r=1$) and its complement perfect quasi-complete graph ($r = n-1$). The normalisation by quasi-star graphs, Fig. \ref{fig1}, centre, shows clear violations of the $[0,1]$ normalisation by the quasi-complete graphs-- as is expected in cases where the quasi-complete graph has larger degree variance than the quasi-star graph-- as well as low densities of thresholded weighted graph models. The normalisation by average degree increases proportionally with $n$ thus is critically flawed as a normalisation. We shall refrain from using it further and focus on the other indices.

All of the graphs show a remarkable stability with respect to network size of small networks, suggesting that this normalisation is very suitable for comparing networks of different sizes and similar density. Furthermore, for non-extremal values of the EEG networks, there is a marked stability with respect to density also, supporting the claim that this normalisation is unbiased to network density. To quantify these statistically, we compute the coefficient of variance (i.e. the ratio of the standard deviation to the mean) of the normalisations with respect to network size for each density, averaging over densities, and the coefficient of variance of the normalisations with respect to density, averaging over network sizes, respectively.

The resulting average coefficients of variation are reported in Table \ref{table1}. Smaller values show less variability, with bold indicating best performance for each network. The proposed normalisation has the least variability with respect to network size for quasi-complete graphs, random geometric graphs and EEG networks. For quasi-star graphs, normalisation by quasi-star achieves a variability of 0, redundantly. For ER random graphs, $\bar{v}$ is second only to $\sigma^{2}$. The result for $\sigma^{2}$ here is quite anomalous, but it is probably achieved actually because it is such a poor normalisation of network size for general networks-- for our normalisation, for instance, ER random graph values are inversely proportional to $n$ \eqref{randnorm}.

With respect to density, the proposed normalisation has the least variability for quasi-complete graphs, ER random graphs and random geometric graphs. Again, for quasi-star graphs, normalisation by quasi-star achieves a variability of 0, redundantly. For EEG networks, $\bar{v}$ is second only to $J$. On inspection of Fig \ref{fig1}, top left and right, however it would appear that this is due to the extremal values. For $\bar{v}$, values fall steeply towards at lowest and highest densities for all network sizes. For $J$ however, some fall steeply towards 0 and others fall steeply upwards, thus perhaps cancelling each other out in the coefficient of variation.

Indeed, this steep fall towards zero is a notable feature in values of $\bar{v}$ of EEG networks. This means that at low densities, the networks are far from star-like, more similar to quasi-complete graphs, whereas at highest densities, the networks are far from quasi-complete, showing properties more similar to quasi-star networks. This can be interpreted in light of the well known ``rich-club'' phenomenon of brain networks-- nodes with lots of connections are connected particularly strongly together \cite{vandenHeuvel2011}. This means that at sparse densities the rich-club evolves as an almost complete subnetwork, keeping heterogeneity low. On the other hand, the dominance of connections to hub nodes means that at very high densities, hub nodes and highly connected nodes become saturated (i.e. share edges with all other nodes), leaving, for want of a better term, a `poor-club' of weakly interconnected nodes, similar to quasi-star graphs.

\begin{table}[t]
	\centering
	\caption{\label{table1} Coefficient of variation of network indices with respect to network size, averaged over density.}
	\begin{tabular}{ccccc}
			Network 			& $\bar{v}$ & $J$ & $\sigma^{2}$ & $\rho$\\
			\hline
			Quasi-star 			& 0.0249			& \textbf{0} 	& 0.7647			& 0.5831	\\
			Quasi-complete 		& \textbf{0.0812} 	& 0.1600		& 0.8512  			& n/a		\\
			ER Random 			& 0.7848			& 0.8949		& \textbf{0.0585}	& 1.0893	\\
			Random Geometric 		& \textbf{0.2240}	& 0.3432		& 0.6418			& 0.4660	\\
			Scale-free	 model 		& \textbf{0.2251}	& 0.2766		& 0.6659			& 0.5211	\\
			Random Hierarchy		& \textbf{0.2454}	& 0.3272		& 0.7091			& 0.4642	\\		
			EEG 				& \textbf{0.0540}	& 0.1825		& 0.7724			& 0.5485	\\

	\end{tabular}
	\bigskip
	%	\footnotetext[5]{And etc.}
	\caption{\label{table15}Coefficient of variation of network indices with respect to density, averaged over network size.}
	\begin{tabular}{ccccc}
			Network 			& $\bar{v}$ & $J$ & $\sigma^{2}$ & $\rho$\\
			\hline
			Quasi-star 			& 0.4320			& \textbf{0	}		& 1.1026		& 1.4377	\\
			Quasi-complete 		& \textbf{0.4509}	& 0.4786			& 0.4877		& n/a 	\\
			ER Random 			& \textbf{0.0378}	& 0.6076			& 0.5793		& 1.0417	\\
			Random Geometric 		& \textbf{0.3756}	& 0.5469			& 0.4606		& 0.6287	\\
			Scale-free model		& \textbf{0.1092}	& 0.1733 			& 0.1480		& 0.5854	\\
			Random Hierarchy		& 0.3135			& \textbf{0.2972}			& 0.6924		& 1.1754	\\	
			EEG 				& 0.2610			& \textbf{0.2386}	& 0.6085		& 0.8387	\\	

	\end{tabular}
\end{table}

\subsection{Computational efficiency of the closed-form expression}
\begin{figure}[t]
	\centering
	\includegraphics[trim= 0 0 0 0,clip,scale = 0.25]{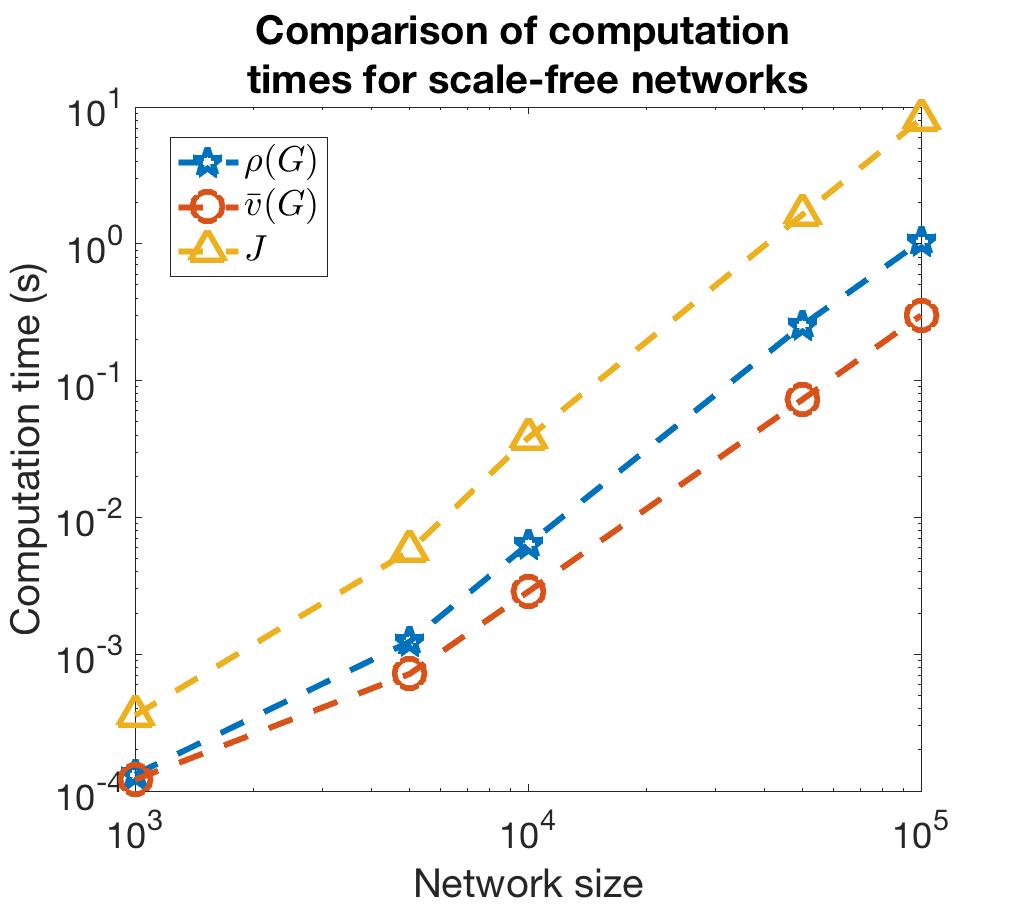}
	\caption{The average computation times of degree variance normalisations $\bar{v}$, $J$ and $\rho$ of scale-free networks of size 5,000, 10,000, 50,000 and 100,000. Both axes are on log scales. The closed form expression of $\bar{v}$ clearly outperforms $J$ which requires construction of a quasi-star graph in the computation. It also outperforms $\rho$ which employs the construction of the graph Laplacian in its computation.}
	\label{CT}
\end{figure}

As the networks grow large, the previously analysed variability becomes negligible for $\bar{v}$ and $J$, where it appears there is a possible asymptotic convergence to a set limiting curve for each network type in Fig. \ref{fig1}. However, what becomes more important as networks grow large is rather the computational cost. The computational efficiency of $\bar{v}$ compared to $J$ can be garnered by comparing processing times of degree variance normalisations for larger graphs. We also compare with $\rho$ for completeness, which uses the graph Laplacian in its computation \cite{Estrada2010}. We use sparse scale-free graphs \cite{Barabasi1999} at 1\% density to demonstrate this. We will look at graphs of size 5,000, 10,000, 50,000 and 100,000. Scale-free graphs of density 1\% can be obtained by selecting 1\% of nodes as the core subgraph and the average degree of additional nodes as $0.5\%n$. The computation time using Matlab algorithms on a single core of a 3.6 GHz Intel Core i7Processor 4274 HE with 32 GB 2400 MHz DDR4 is computed 25 times and the average time is shown in Fig. \ref{CT}.

This clearly demonstrates the increased computational efficiency of $\bar{v}$ over $J$ as a normalisation of degree variance. Indeed, $\bar{v}$ was computed an order of 10 times faster at each network size and the trend appears to show that this difference grows with greater network sizes. For example, $\bar{v}$ was computed for 10,000 node networks with an average speed of 0.003s while $J$ was computed with an average speed of 0.038s. Scaling to 100,000 node networks the respective average speeds were 0.300s and 8.163s-- a factor of over 27 in difference. Furthermore, $\bar{v}$ has greater computational efficiency than $\rho$ which showed average speeds of 0.006 for 10,000 nodes and 1.043 for 100,000 nodes, the latter being a factor of roughly 3.5 in difference.

\subsection{Normalised degree variance of real-world binary networks}
\begin{figure*}[t]
	\centering
	\includegraphics[trim= 0 0 0 0,clip,scale = 0.175]{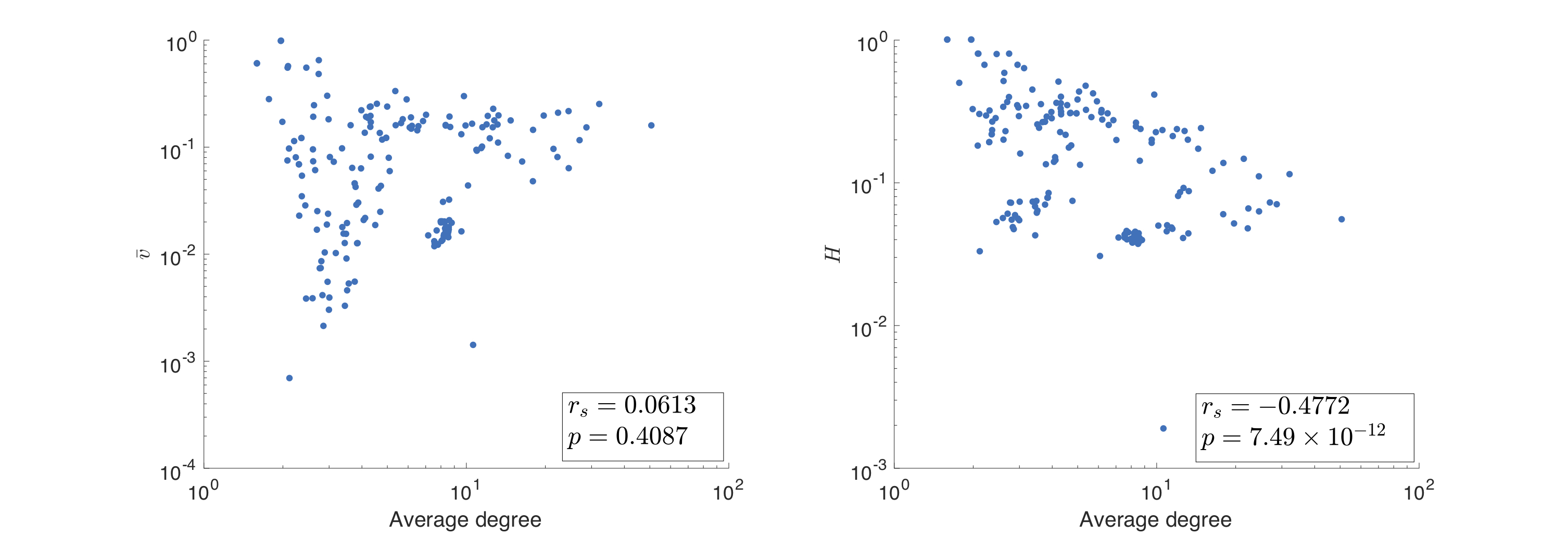}
	\caption{Normalised degree variance, left, and heterogeneity, right, of 184 real-world networks plotted against their average degrees.}
	\label{ICON}
\end{figure*}

We analysed the 184 static networks taken from the Colorado Index of Complex Networks (ICON) \cite{Clauset2016} as used in \cite{Ghasemian2018}, covering a variety of biological, social and technological networks. Using the rank-based Spearman correlation coefficient ($r_{s}$), network size was found to be strongly anti-correlated with density in these networks ($r_{s} = -0.840, p = 3.910\times10^{-50}$). For this reason, and since did not have large enough samples at any specific size or density, we assessed Spearman correlations ($r_s$) between normalised heterogeneity indices of these networks and their average degrees, $(n-1)d$. The values for $\bar{v}$ and $\rho$ are plotted against average degree in Fig \ref{ICON}. There was a clear trend noted with $\rho$ and average degree, Fig \ref{ICON}.  Normalised degree variance and average degree showed a small positive correlation ($r_{s} = 0.0210, p = 0.1701$). However there was a stronger negative correlation between $\rho$ and average degree ($r_{s} = -0.2899, p = 6.56\times10^{-5}$). The median and interquartiles (25th and 75th percentiles) of the value of normalised degree variance for real world networks was $0.1203 \ [0.0245,0.1904]$, while for the normalised heterogeneity index it was $0.2350 \ [0.0890,0.3437]$. It was observable, however, that the range of values obtained for both metrics was lower at higher average degrees. This is clear from Figure \ref{ICON}, left, where for average degrees of between 2 and 4 the normalised degree variance values are spread out, taking values between 1 and 0.001. On the other hand, networks with average degree upwards of 4 show a smaller spread, mostly concentrated betewen 0.3 and 0.04. A similar behaviour is seen for network heterogeneity, Figure \ref{ICON}, right.

\begin{figure*}[t]
	\centering
	\includegraphics[trim= 0 0 0 0,clip,scale = 0.2]{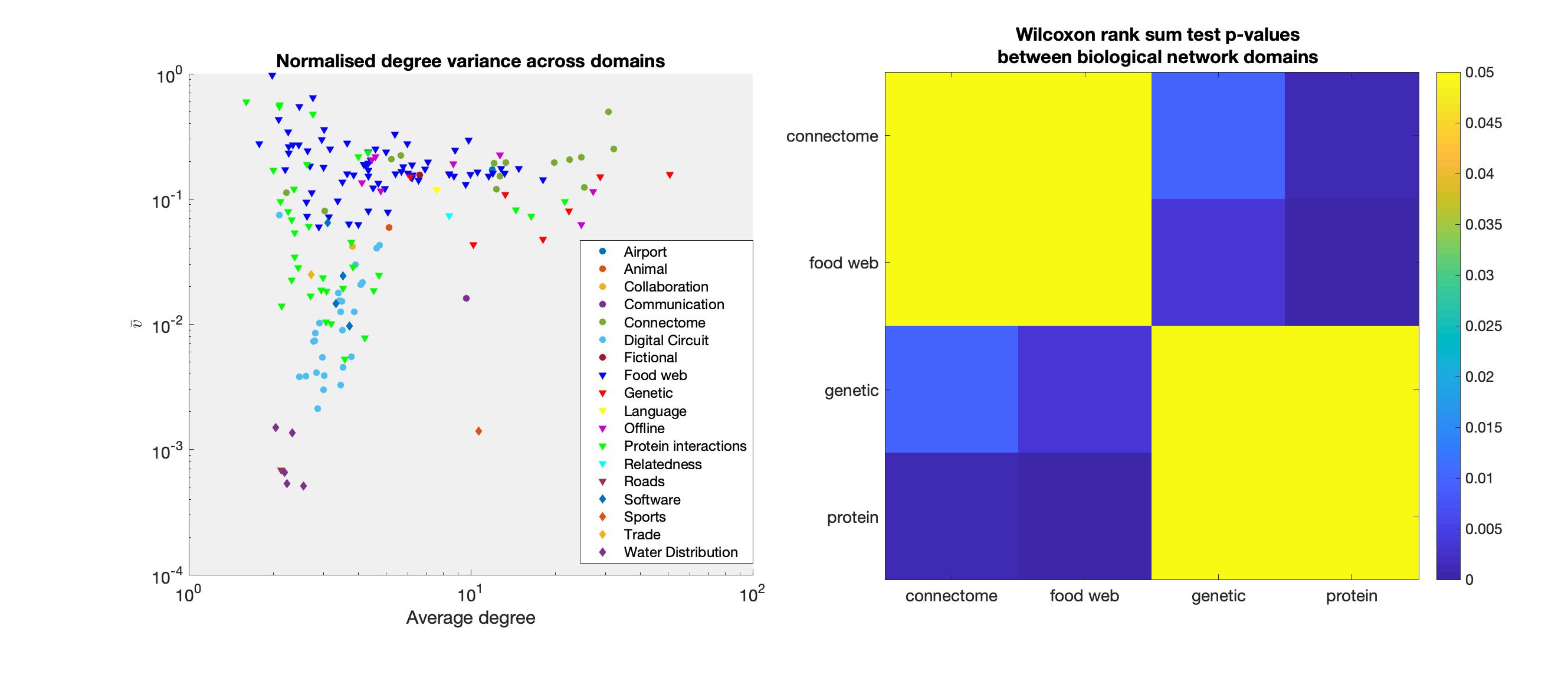}
	\caption{Normalised degree variance of 184 real-world networks plotted against their average degrees and marked by sub domain, left, and Wilcoxon rank sum tests of normalised degree variance within biological network domains, right.}
	\label{dom}
\end{figure*}

From this dataset, we can see how the greater independence of the proposed normalisation aids in understanding general trends of degree variance across domains, Fig \ref{dom}, left. It is evident that biological/animal networks such food webs, animal connectomes and protein networks show greater degree distribution heterogeinety than technological and infrastructural networks, such as water distribution networks, road networks and digital circuit networks. Furthermore, four biological network subdomains (animal brain connectomes, food webs, genetic networks and protein interaction networks) have multiple samples (14, 71, 7 and 38, respectively) to allow assessment of group differences. These were tested using Wilcoxon's rank sum test, and the results are as in Fig \ref{dom}, right. Interestingly, animal connectomes and food webs were comparable in terms of degree distribution heterogeneity (bright yellow indicates no difference with $p>0.05$) while these were statistically larger in general than genetic networks and protein interaction networks. This suggests that food webs and connectomes have a stronger hierarchical structure than genetic and protein interaction networks. Indeed, connectomes are known to have a rich-club of cognitive processing regions which integrate information across lower-order processes such as sensorimotor and heteromodal regions \cite{Smith2019}, while such a rich-club with wide-ranging integrative processes is not seen at the genetic/protein level \cite{Maslov2002}.

For each real-world network from the ICON corpus, we randomly took out different percentages (5\%, 10\%, 15\%, 20\% and 25\%) of nodes and edges from the networks and computed $\bar{v}$ and $\rho$ of the resulting subnetworks. This was done 50 times for each percentage and the mean was taken as an approximation of the expected value for that percentage. Nodes were removed uniformly at random in each iteration. Edges were removed in two ways-- uniformly at random and randomly with probabilities inversely proportional to the product of the degrees of their adjacent nodes (i.e. probability of removal $p_{r}\sim((n-1)-k_{i})((n-1)-k_{j})$), the latter of which better retains the expected topology of the network from an evolutionary perspective. Specifically, this was done by i) computing $h_{ij} = ((n-1)-k_{i})((n-1)-k_{j})$ for all edges $ij$, ii) computing a vector of the cumulative sum of these values, iii) dividing these values by the sum total, $T = \sum_{i,j}h_{ij}$, to normalise the vector in [0,1], iv) sampling the space [0,1] uniformly at random such that wherever the value fell with respect to the normalised cumulative vector corresponded to the edge that was subsequently removed from the network. That is, for $m$ edges, the interval [0,1] was split into $m$ bins whose sizes were proportional to the $h_{ij}$'s such that the probability that a random sample of [0,1] fell in the bin proportional to $h_{ij}$ was exactly $h_{ij}/T$.

In each subsampling approach, the resulting subnetworks often contained isolated nodes leading to failure in the computation of $\rho$. This was the case for an average of 36.3\% of the time for a given real world network with nodes removed and 65.1\% and 67.1\% of the time for edges removed by degree and randomly, respectively. This highlights the intrinsic problem of the $\rho$ metric for broad use in network studies. Still, these were disregarded and the mean value was taken only for those iterations which produced a subnetwork with no isolated nodes. The median absolute difference across the 184 real world networks was then computed for each metric at each percentage of nodes and edges removed. The results are as in Table \ref{ICONtab} where $\bar{v}_{sub}$ and $\rho_{sub}$ denote the corresponding metric values of the subnetworks. It is quite clear that, on average, the difference found for $\bar{v}$ is generally less than found for $\rho$ for both node and edge subsampling. Moreover the differences are clearly fairly small,  with the worst case scenario presented (25\% of edges randomly removed) showing a median change in $\bar{v}$ of just less than 0.01, indicating that the index is indeed robust to subsampling. This robustness was found to be stronger for node subsampling than for edge subsampling as seen by the fact that the median change in $\bar{v}$ was generally smaller in node subsampling than in edge subsampling for the same percentage removed. This is compounded by the fact that for a given number of nodes removed, we found that generally a higher percentage of edges were removed as a consequence, Table \ref{ICONtab}, row 4. Further, as expected, $\bar{v}$ was more robust in the edge subsampling approach following the evolutionary considerations than it was for the uniformly random subsampling approach, whereas, surprisingly the opposite was found for $\rho$ which may be indicative of the restriction of results for $\rho$ within subsamples which found no isolated nodes.

\begin{table}[t]
	\centering
	\caption{\label{ICONtab} Median robustness of normalised degree variance, $\bar{v}$, and heterogeneity, $\rho$, to subsampling across 184 real-world networks}
	\begin{tabular}{c|ccccc}
			\textbf{\% $n$	removed}		& 5\% 	& 10\% 	& 15\% 	& 20\% 	& 25\%\\
			\hline
			med$(|\bar{v}-\bar{v}_{sub}|)$ 		&0.0004 	&0.0009 	&0.0013 	&0.0017 	&0.0022	\\
			med$(|\rho-\rho_{sub}|)$ 			&0.0013 	&0.0024 	&0.0034 	&0.0034 	&0.0039	\\
			\% $m$ lost (mean)				&9.7\% 	&19.6\% 	&28.1\% 	&36.3\% 	&44.3\%	\\
			\hline
			\textbf{\% $m$ removed}		& 5\% 	& 10\% 	& 15\% 	& 20\% 	& 25\%\\
			\textbf{inversely to degree}\\
			\hline
			med$(|\bar{v}-\bar{v}_{sub}|)$ 		&0.0012 	&0.0023 	&0.0034 	&0.0043 	&0.0050	\\
			med$(|\rho-\rho_{sub}|)$ 			&0.0027 	&0.0052 	&0.0079 	&0.0107 	&0.0131	\\
			\hline
			\textbf{\% $m$ removed}		& 5\% 	& 10\% 	& 15\% 	& 20\% 	& 25\%\\
			\textbf{randomly}\\
			\hline
			med$(|\bar{v}-\bar{v}_{sub}|)$ 		&0.0018 	&0.0040 	&0.0067 	&0.0078 	&0.0099	\\
			med$(|\rho-\rho_{sub}|)$ 			&0.0022 	&0.0045 	&0.0069 	&0.0101 	&0.0125	\\
	\end{tabular}
\end{table}

\section{Conclusion}
We introduced and mathematically justified a true normalisation for degree variance in networks, showing a large degree of invariance to network size and network density. The two other previously proposed normalisations were shown to be invalid and another normalised heterogeneity index was shown to be ill-defined for graphs with isolated nodes. Beyond this, our normalisation had generally lower variability with respect to network size in quasi-complete graphs, ER random graphs, random geometric graphs, scale-free networks, random hierarchy models and EEG networks. Using sparse scale-free models, it was also shown to be more computationally efficient than other normalisation approaches. We then demonstrated the normalisation's applicability and robustness to subsampling in real-world networks. The usefulness of the closed form expression for normalisation was also demonstrated in the production of two mathematical results showing that normalised degree variance of quasi-star graphs decreases with respect to density and that a flexible lower bound for normalised degree variance was possible depending on proportions of given degrees within the network. All things considered, the proposed normalisation is put forward as a standard for measuring heterogeneity in complex networks.

\section{List of Abbreviations}
ER- Erd\"os-R\'enyi; EEG- Electroencephalogram; ICON- Colorado Index of Complex Networks
\section{Declarations}
\subsection{Availability of data and materials}
The code for normalised degree variance and the experiments conducted is available permanently on the Open Science Framework:  DOI 10.17605/OSF.IO/2VA7Y. The real network dataset analysed during the current study is available in the Colorado Index of Complex Networks repository, \url{https://icon.colorado.edu} \cite{Clauset2016}.
\subsection{Competing interests}
The authors declare that they have no competing interests.
\subsection{Funding}
KS was supported by Health Data Research UK (MRC ref Mr/S004122/1), which is funded by the UK Medical Research Council, Engineering and Physical Sciences Research Council, Economic and Social Research Council, National Institute for Health Research (England), Chief Scientist Office of the Scottish Government Health and Social Care Directorates, Health and Social Care Research and Development Division (Welsh Government), Public Health Agency (Northern Ireland), British Heart Foundation and Wellcome.
\subsection{Author's contributions}
KS and JE conceived the study.  KS performed the analyses and wrote the manuscript. JE reviewed the manuscript. All authors read and approved the final manuscript.
\subsection{Acknowledgements}
Not applicable.

\section*{Appendix I: Normalised degree variance of $G(n,p)$}
Complementary to Corollary \ref{Gnm} for $G(n,m)$, the following is a proof that the normalised degree variance of ER random graph ensemble $G(n,p)$ does not depend on the number of expected edges $m$, and thus on network density, for fixed $n$.
\begin{corollary}
For the ER random graph ensemble $G(n,p)$
\begin{equation}
E[\bar{v}(G(n,q))] \approx \frac{2(n-1)}{n^2}.
\end{equation}
\begin{proof}
Recall that, generally, $G(n,p)$ has a binomial degree distribution, $B(n,p)$. The variance of this distribution is then $(n-1)p(1-p)$. This gives
\begingroup
\addtolength{\jot}{0.75em}
\begin{align}
E[\bar{v}(G(n,p))] 	&= \frac{n-1}{nm(1-d)}(n-1)p(1-p)\\
				&= \frac{2(n-1)}{n^2}\frac{p(1-p)}{d(1-d)},
\end{align}
\endgroup
recalling that $d = 2m/n(n-1)$. For very large $n$, it is almost guaranteed that $m = \frac{n(n-1)}{2}p$ giving $d = p$ and in such case it is evident that the above simplifies to $2(n-1)/n^2$. However, more precisely, $m$ is determined by $n(n-1)/2$ Bernoulli trials with probability $p$ and so is distributed binomially as $B(n(n-1)/2,p)$. Therefore, writing $M = n(n-1)/2$, and $C = 2(n-1)/n^2$, we have
\begingroup
\addtolength{\jot}{0.75em}
\begin{align}
E[\bar{v}(G(n,p))] 	&= Cp(1-p)E\left[\frac{1}{d(1-d)}\right]\\
				&= Cp(1-p)E\left[\frac{1}{\frac{m}{M}(1-\frac{m}{M})}\right]\\
				&=^{*} Cp(1-p)\sum_{m=0}^{M}\tfrac{1}{\tfrac{m}{M}\left(1-\tfrac{m}{M}\right)} {M \choose m}p^{m}(1-p)^{M-m} \\ 
				&= C\sum_{m=0}^{M}\frac{M^2}{m(M-m)}\frac{M!}{m!(M-m)!}p^{m+1}(1-p)^{M-m+1},\label{eq64}
\end{align}
\endgroup
where $^{*}$ comes from the law of the unconscious statistician. Now, $m= 0$ and $m = M$ are the special cases of the empty and complete graphs, respectively. These put a 0 on the denominator, but because these are two extremely unlikely cases for any relevant $p$ (having the smallest binomial coefficients), they can be discarded without loss of accuracy in approximation. Now, from the binomial formula, we have 
\begin{align}
\sum_{m=0}^{M+2} {M+2\choose m}p^{m}(1-p)^{M+2-m} = (p + 1-p)^{M+2} = 1,
\end{align}
but, again, we can approximate this by discarding negligible terms at the periphery: $m=0,1,M, M+1$ and $M+2$, i.e.
\begin{align}
\sum_{m=1}^{M-1} {M+2\choose m+1}p^{m+1}(1-p)^{M+2-(m+1)} \approx 1.
\end{align}
By simple algebraic manipulations, we can then rearrange equation \eqref{eq64} to get
\begingroup
\addtolength{\jot}{0.75em}
\begin{align}
	C\sum_{m=1}^{M-1}\frac{(m+1)(M-m+1)M^2}{m(M+1)(M+2)(M-m)}{M+2\choose m+1}p^{m+1}(1-p)^{M+2-(m+1)}.\\
\end{align}
\endgroup
It then remains to show that the terms
\begin{equation}\label{uglycoeff}
B_m = \frac{(m+1)(M-m+1)M^2}{m(M+1)(M+2)(M-m)}
\end{equation}
approximate unity. Now $M>m$ and $m$ ranges from 1 to $M-1$, but also the binomial coefficients are dominated by the middle terms, so we must consider the leading terms of equation \eqref{uglycoeff} when $m \approx M/2$. In this case
\begingroup
\addtolength{\jot}{0.75em}
\begin{align}
B_{\frac{M}{2}} 	&= \frac{(M/2 + 1)^2M^2}{(M/2)^2(M+1)(M+2)}\\
			&= \frac{M^2 + 4M+4}{M^2+3M+2}\approx 1
\end{align}
\endgroup
for large $M$, as required. So, finally, we have
\begingroup
\addtolength{\jot}{0.75em}
\begin{align}
E[\bar{v}(G(n,p))] &\approx C\sum_{m=1}^{M-1}B_{m}{M+2 \choose m+1}p^{m+1}(1-p)^{M+2-(m+1)}\\
	&\approx C\sum_{m=1}^{M-1}{M+2 \choose m+1}p^{m+1}(1-p)^{M+2-(m+1)}\\
	&\approx C = \frac{2(n-1)}{n^{2}},
\end{align}
as required.
\endgroup
\end{proof}
\end{corollary}

\section*{Appendix II: Network size-independent minimum of normalised degree variance for graphs with a fixed proportion of vertices of a given degree}
%\subsection*{}
Here we consider lower bounds for the normalisation's invariance to network size. Suppose for a graph, $G$, we guarantee $x$ nodes of degree $a$, having degrees $\mathbf{k}=\{a,a,\dots,a,k_{x+1},\dots,k_{n}\}$. Then 
\begingroup
\addtolength{\jot}{0.75em}
\begin{align}
\bar{v}(G) 	&= \frac{1}{1-d}\left(\frac{\sum_{i=1}^{n}k_{i}^{2}}{nm} - \frac{4m}{n^{2}}\right)\\
		&= \frac{1}{1-d}\left(\frac{xa^2 + \sum_{i=x+1}^{n}k_{i}^{2}}{nm}- \frac{4m}{n^{2}}\right)\\
		&\geq^{*} \frac{1}{1-d}\left(\frac{xa^2 + (n-x)\left(\frac{2m-xa}{n-x}\right)^{2}}{nm}- \frac{4m}{n^{2}}\right)\\
%&= \tfrac{1}{1-d}\left(\tfrac{xa^2n^2-a^2x^2n + 4m^2n -4axmn + a^2x^2n-4m^2n+4xm^2}{n^{2}(n-x)m}\right)\\
		& = \frac{1}{1-d}\left(\frac{x(a^2n^{2} - 4amn + 4m^{2}) }{n^{2}(n-x)m}\right)\\
		& = \frac{1}{1-d}\left(\frac{x(an - 2m)^{2}}{n^{2}(n-x)m}\right),
\end{align}
\endgroup
where * comes from the fact that $\sum_{i=1}^{n}k_{i} = 2m$ and that the minimum value of the sum of squares come from all elements being equal $\implies k_{x+i} = (2m - x)/(n-x) \ \forall i$.

Now, we fix density so that $d = 2m/n(n-1)$, i.e. $2m = dn(n-1)$ and we get 
\begin{equation}\label{eqAppB}
\frac{2x(an - dn(n-1))^{2}}{d(1-d)n^{3}(n-1)(n-x)}.
\end{equation}

Then, \eqref{eqAppB} = 0 
\begingroup
\addtolength{\jot}{0.75em}
\begin{align}
&\iff an = dn(n-1)\\
&\iff \frac{a}{n-1} = d = \frac{2m}{n(n-1)}\\
&\iff a = \frac{2m}{n}
\end{align}
\endgroup
which is the average degree of the graph, which makes sense since regular graphs have degree variance $0$ where each node has degree $2m/n$. Taking $a$ to be small, say 1, the leading terms in \eqref{eqAppB} are
\begin{equation}\label{eqAppB2}
\frac{2xd^2n^4}{d(1-d)n^5} = \frac{x}{n}\frac{2d}{1-d},
\end{equation}
Noting that $x/n$ is the proportion of nodes which have degree 1, we thus have a constant minimum value of degree variance for such graphs with known density $\forall n$.

We can also consider $a=n-1$, i.e. graphs with a known proportion of dominant vertices. Then, \eqref{eqAppB} becomes
\begin{equation}\label{eqAppB3}
\frac{2x((1-d)n(n-1))^2}{d(1-d)n^3(n-1)(n-x)} \approx \frac{x}{n}\frac{2(1-d)}{d},
\end{equation}
considering the leading terms. Notably, this is precisely \eqref{eqAppB2} if we substitute in $d$ for $1-d$, reflecting the equivalence of $\bar{v}$ for graphs and their complements.

\begin{figure}[!tb]
	\centering
	\includegraphics[trim= 0 0 0 0,clip,scale = 0.25]{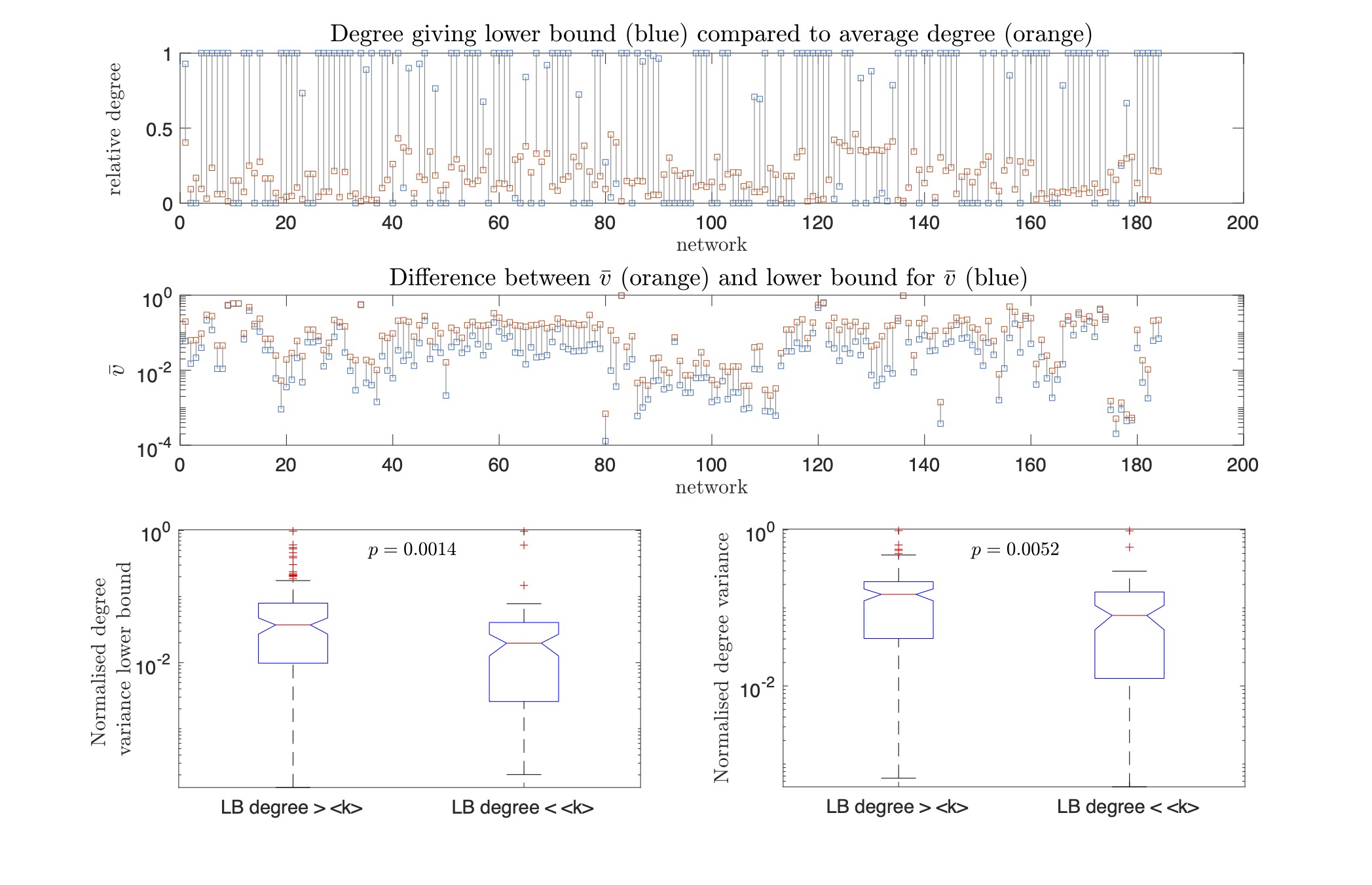}
	\caption{Analysis of the lower bounds for 184 real world networks. The $p$-values in the box plots come from Wilcoxon rank sum tests. Networks whose lower bound was achieved at a degree larger than the average degree are compared against networks whose lower bounds are at degrees lower than the average. These are recorded for the lower bounds for normalised degree variance (left) and normalised degree variance (right). LB degree is the degree at which the lower bound is achieved in the network and $<k>$ is the average degree of the network.}
	\label{LowerBound}
\end{figure}

Therefore, if the proportions of a given degree of a network is known, then a lower bound of $V$ for such networks can be established. This could help, for example, in questions of hub dominance in real world networks. One can fix the number of dominant nodes in the network and know the range of $V$ possible for such networks with greater accuracy. %We shall now go on to apply our normalisation to network models and real world data to show its relevance to problems of network comparability and computational efficiency. 
We applied these results to investigate lower bounds of $\bar{v}$ for the 184 real world networks. To get the maximum lower bound for a single network, equation \eqref{eqAppB} was computed for all degrees and the maximum of these alongside the degree at which the maximum was obtained were recorded. To help comparisons between different networks, the degrees were recorded as relative degrees:
\begin{equation}
	\bar{k}_{i} = \frac{k_{i} - min(\mathbf{k})}{max(\mathbf{k})-min(\mathbf{k})},
\end{equation}
so that the maximum is 1 and minimum is 0.  The results are plotted in Fig \ref{LowerBound}. Most lower bounds are achieved at either the maximum degree (96 networks) or minimum degree (60 networks) in the network, Fig \ref{LowerBound} top. The achieved value for degree variance was on average (median) 3.33 times the value of the lower bound, with an interquartile range of 2.62. Absolute differences between degree variance and lower bounds are shown in Fig \ref{LowerBound} row 2, on a log scale. Using Wilcoxon rank sum tests, we found that networks whose lower bounds were achieved at degrees larger than the network's average degree generally had greater normalised degree variance ($p = 0.0052$) and lower bounds ($p = 0.0014$) than those networks whose lower bounds were achieved at degrees lower than the network's average degree, Fig \ref{LowerBound} bottom. These results all align and support the claim that degree variance provides a measurement of the inequality of the degree distribution.

\end{document}